\documentclass[prb,reprint,superscriptaddress]{revtex4-2}

\usepackage[utf8]{inputenc}
\usepackage{blindtext}
\usepackage{svg}
\usepackage{subfiles}
\usepackage{amsmath}
\usepackage{amsfonts}
\usepackage{hyperref}
\usepackage{wasysym}
\usepackage{bbold}
\usepackage{comment}

\renewcommand{\[}{\begin{equation}}
\renewcommand{\]}{\end{equation}}
\newcommand{\ket}[1]{\left|#1\right>}
\newcommand{\bra}[1]{\left<#1\right|}

\newcommand{\dth}{\Delta\theta}
\newcommand{\sinc}{\textrm{sinc}}

\usepackage{upgreek}
\newcommand\ueV{{\upmu\textrm{eV}}}

\newcommand\us{{\upmu\textrm{s}}}

\newcommand\tf{{t_\textrm{f}}}
\newcommand\Hz{\textrm{Hz}}
\newcommand\MHz{\textrm{MHz}}

\newcommand\kB{{k_\textrm{B}}}

\newcommand\xct[1]{{\langle #1 \rangle}}

\newcommand\inv{^{-1}}
\newcommand\half{\frac{1}{2}}

\newcommand\tmin{\textrm{min}}
\newcommand\tmax{\textrm{max}}

\newcommand\round[1]{ { \left(        #1 \right)       }}

\renewcommand\tf{{t_\textrm{f}}}
\renewcommand\tt{{T_2^*}}
\newcommand{\Dt}{{\Delta t}}

\newcommand{\omegaLow}{\omega_\textrm{low}}
\newcommand{\Dom}{{\Delta\omega}}

\graphicspath{ {images/} }

\makeatletter
\DeclareFontFamily{OMX}{MnSymbolE}{}
\DeclareSymbolFont{MnLargeSymbols}{OMX}{MnSymbolE}{m}{n}
\SetSymbolFont{MnLargeSymbols}{bold}{OMX}{MnSymbolE}{b}{n}
\DeclareFontShape{OMX}{MnSymbolE}{m}{n}{
    <-6>  MnSymbolE5
   <6-7>  MnSymbolE6
   <7-8>  MnSymbolE7
   <8-9>  MnSymbolE8
   <9-10> MnSymbolE9
  <10-12> MnSymbolE10
  <12->   MnSymbolE12
}{}
\DeclareFontShape{OMX}{MnSymbolE}{b}{n}{
    <-6>  MnSymbolE-Bold5
   <6-7>  MnSymbolE-Bold6
   <7-8>  MnSymbolE-Bold7
   <8-9>  MnSymbolE-Bold8
   <9-10> MnSymbolE-Bold9
  <10-12> MnSymbolE-Bold10
  <12->   MnSymbolE-Bold12
}{}

\let\llangle\@undefined
\let\rrangle\@undefined
\DeclareMathDelimiter{\llangle}{\mathopen}%
                     {MnLargeSymbols}{'164}{MnLargeSymbols}{'164}
\DeclareMathDelimiter{\rrangle}{\mathclose}%
                     {MnLargeSymbols}{'171}{MnLargeSymbols}{'171}
\makeatother

\begin{document}

\author{F.~Fehse}
\affiliation{Department of Physics, McGill University, 3600 rue University, Montr\'eal, QC H3A 2T8, Canada}
\author{M.~David}
\affiliation{Department of Physics, McGill University, 3600 rue University, Montr\'eal, QC H3A 2T8, Canada}
\affiliation{D\'epartement de Physique de l'\'Ecole Normale Sup\'erieure, PSL Research University, 75005 Paris, France}
\author{M.~Pioro-Ladri\`ere}
\affiliation{D\'epartement de Physique, Universit\'e de Sherbrooke, Sherbrooke, QC J1K 2R1, Canada}
\affiliation{Institut Quantique, Université de Sherbrooke, Sherbrooke, QC J1K 2R1, Canada}
\author{W.~A.~Coish}
\affiliation{Department of Physics, McGill University, 3600 rue University, Montr\'eal, QC H3A 2T8, Canada}

\title{Generalized fast quasiadiabatic population transfer for improved qubit readout, shuttling, and noise mitigation}

\date{\today}

\begin{abstract}
Population-transfer schemes are commonly used to convert information robustly stored in some quantum system for manipulation and memory into more macroscopic degrees of freedom for measurement. These schemes may include, e.g., spin-to-charge conversion for spins in quantum dots, detuning of charge qubits between a noise-insensitive operating point and a measurement point, spatial shuttling of qubits encoded in spins or ions, and parity-to-charge conversion schemes for qubits based on Majorana zero modes. A common strategy is to use a slow (adiabatic) conversion. However, in an adiabatic scheme, the adiabaticity conditions, on the one hand, and accumulation of errors through dephasing, leakage, and energy relaxation processes on the other hand, limit the fidelity that can be achieved. Here, we give explicit fast quasiadiabatic (fast-QUAD) conversion strategies (pulse shapes) beyond the adiabatic approximation that allow for optimal state conversion. In contrast with many other approaches, here we account for noise in combination with pulse shaping. Although we restrict to noise sources that can be modeled by a classical fluctuating parameter, we allow generally for anisotropic nonGaussian noise that is nevertheless sufficiently weak to lead to a small error. Inspired by analytic methods that have been developed for dynamical decoupling theory, we provide a general framework for unique noise mitigation strategies that can be tailored to the system and environment of interest. 
\end{abstract}

\maketitle
\section{Introduction}

A common problem in quantum dynamics is the transfer of information from one system to another. For coherent manipulation, it may be useful to encode information in a quantum system having long coherence and relaxation times (such a system may be composed of, e.g., noise-resistant electron-spin, hyperfine, or topologically protected orbital levels). However, to perform a high-quality measurement, it is typically necessary to robustly convert information into more macroscopic degrees of freedom (e.g., charge or photons) that can be easily distinguished by a measurement apparatus.  Specific applications of this information transfer include, e.g., spin-to-charge conversion for spins in quantum dots \citep{johnson2005triplet,maune2012coherent,harveyCollard2018high}, parity-to-charge conversion for qubits based on Majorana zero modes \citep{flensberg2011non,gharavi2016readout,szechenyi2020parity}, shuttling of electron spins \cite{kandel2019coherent,yoneda2021coherent} or ions \cite{kaushal2020shuttling}, storage and retrieval of information in a quantum memory \cite{chaneliere2005storage,lvovsky2009optical,hedges2010efficient}, and conversion between stationary and flying qubits \cite{yao2005theory}.  Typically, this type of information conversion requires a population transfer from the energy eigenbasis of one Hamiltonian to the eigenbasis of another.  

The most common and natural strategy to perform a population transfer is to design a time-dependent Hamiltonian that \emph{adiabatically} interpolates between two eigenbases (typically describing distinct degrees of freedom). However, any strategy relying on adiabaticity must necessarily be slow relative to some time scale. This limits the speed of possible measurements that may be otherwise used for rapid feed-forward processing and error correction in a quantum processor. Slow conversion may also lead to the accumulation of errors through leakage and energy relaxation processes due to uncontrolled terms in the Hamiltonian. There are many alternative strategies (beyond adiabatic conversion) that can be used to accelerate population transfer. These include transitionless quantum driving \cite{berry2009transitionless}, strategies that exploit dressed states \citep{baksic2016speeding}, superadiabatic conversion \citep{zhou2017accelerated, agundez2017superadiabatic}, and counterdiabatic driving \citep{sels2017minimizing}. See also Ref.~\onlinecite{gueryOdelin2019shortcuts} for a review.

In addition to controlling nonadiabatic errors, it is also important to consider errors due to dephasing and dissipation during a dynamical population transfer \cite{kayanuma1985stochastic, gefen1987zener, ao1991quantum, shimshoni1993dephasing,wubs2006gauging,saito2007dissipative,lacour2007uniform,nalbach2009landau,javanbakht2015dissipative,nalbach2014adiabatic,malla2017suppression,krzywda2020adiabatic}. This is especially true in the applications listed above related to measurement, where a common element is that dephasing is minimized at an operating point (e.g., for the `spin' or `parity' quantum number), but measurement-induced dephasing is maximized at the measurement point. Such a quantum system undergoing a population transfer will therefore be exposed to a dephasing process of varying severity during the transfer and it is important to account for this dephasing in a complete study of the error budget. There have been many previous works that account for dephasing and dissipation in population transfer through an avoided crossing during a standard ($t$-linear) Landau-Zener sweep \cite{kayanuma1985stochastic, gefen1987zener, ao1991quantum, shimshoni1993dephasing,wubs2006gauging,saito2007dissipative,lacour2007uniform,nalbach2009landau,javanbakht2015dissipative,nalbach2014adiabatic,malla2017suppression,krzywda2020adiabatic}, but relatively few that account for the interplay of dephasing and dissipation in combination with a nontrivial pulse shape \cite{ban2014counter,blattmann2015qubit,guo2021efficient}. Here, we consider this interplay and show that notions from dynamical decoupling theory can be used to improve a dynamical population transfer when allowing for more general pulse shapes (designed to minimize nonadiabatic errors). Our primary goal in this work will be to use these strategies to improve \emph{population}-transfer schemes. These schemes are sensitive to dephasing during the transfer, but they will generally not require phase coherence between energy eigenstates at the end of the transfer process, so the requirements may not be as strict as for coherent \emph{state} transfer. Population transfer is the typical process required for a qubit readout or for the preservation of ancilla qubits while shuttling. The ideas presented here can nevertheless also be used to improve coherent state transfer for a suitably modified fidelity metric that accounts for phase coherence at the final time.

\begin{figure}
\begin{centering}
\includegraphics[width=1\columnwidth]{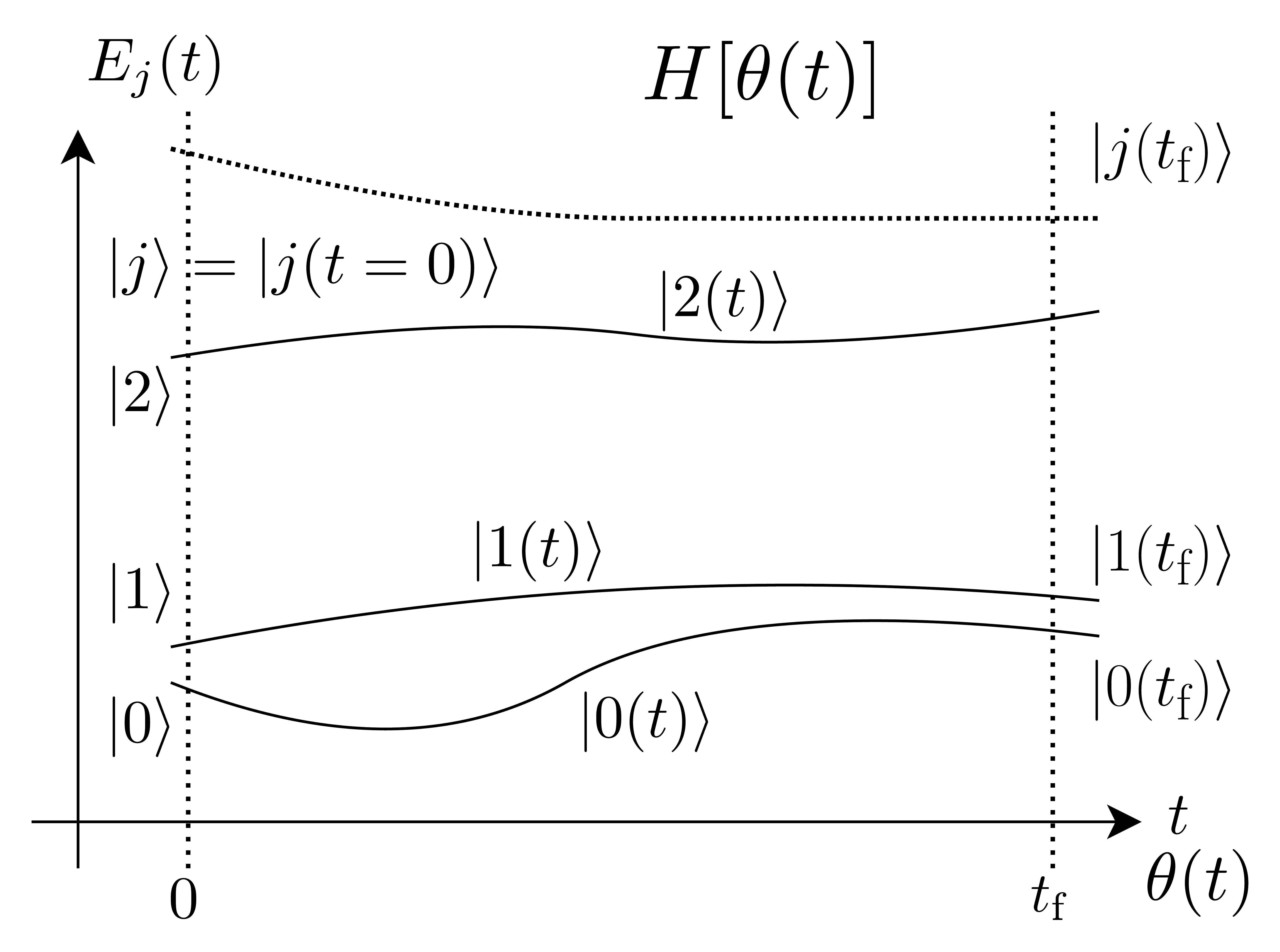}
\par\end{centering}
\caption{Energy levels $E_j(t)$ corresponding to instantaneous eigenstates $|j(t)\rangle$ for a time-dependent Hamiltonian $H[\theta(t)]$, with time $t\in [0,t_\textrm{f}]$. A fast quasiadiabatic (fast-QUAD) pulse offers optimal population transfer in a two-level subspace spanned by $\ket{0(t)}$, $\ket{1(t)}$.}
\label{fig: energy levels}
\end{figure}

\begin{figure}
\begin{centering}
\includegraphics[width=0.8\columnwidth]{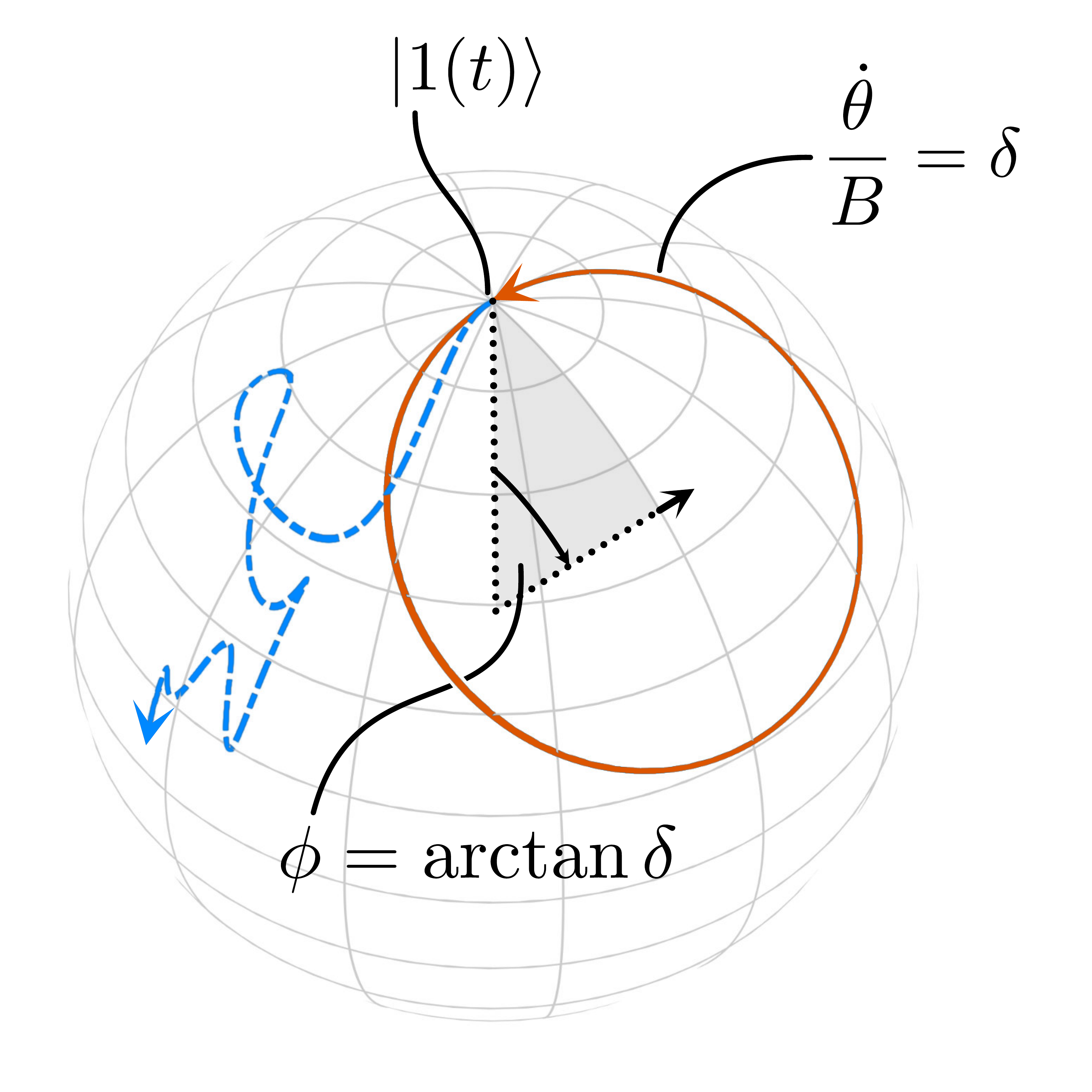}
\par\end{centering}
\caption{\label{fig: trajectory illustration}Possible trajectories of the Bloch vector in the adiabatic frame (where the instantaneous energy eigenstates are located at the north and south poles of the Bloch sphere).  The trajectory for an unoptimized pulse is shown as a dashed blue line. A circular trajectory resulting from a fast-QUAD pulse [a solution to Eq.~(\ref{eq: diff eq})] is shown as a solid orange-red line, with central axis determined by an azimuthal angle $\phi$.}
\end{figure}

For concreteness, we now consider a general population-transfer scheme for a qubit described by a time-dependent Hamiltonian $H(t)$ with instantaneous eigenstates $\ket{j(t)}$ and eigenvalues $E_j(t)$ (see Fig.~\ref{fig: energy levels}).  The goal of population transfer is to map the population of an initial qubit state $\ket{0}$ (or $\ket{1}$) onto the corresponding final qubit state $\ket{0(\tf)}$ (or $\ket{1(\tf)}$), often with the aim of performing a measurement on the final state. In addition to the schemes listed above, a promising strategy for rapid population transfer is provided by a fast quasiadiabatic (fast-QUAD) pulse. For a two-level single-parameter Hamiltonian, $H[\theta(t)]$, with energies $E_j[\theta(t)]$, a fast-QUAD control pulse $\theta(t)$ is given as the solution to the differential equation (setting $\hbar=1$) \cite{roland2002quantum,daems2008quantum,martinezGaraot2015fast,xu2019improving}
\begin{equation}
\frac{\delta}{2}=\frac{\bra{0[\theta(t)]}\frac{d}{dt}\ket{1[\theta(t)]}}{E_1[\theta(t)]-E_0[\theta(t)]}=\frac{\dot{\theta}(t)}{2B(t)}=\mathrm{const.}.
\label{eq: diff eq}
\end{equation}
Here, we have introduced the time-dependent level splitting, $B(t)=E_1(t)-E_0(t)$, and the parameter $\delta$ is taken to be constant (time-independent) for the duration of the pulse. The usual adiabatic regime is realized for $\delta\ll 1$. In this regime, the populations are transferred with low error (a state that starts near the north pole in Fig.~\ref{fig: trajectory illustration} will not deviate significantly). In contrast, a large constant $\delta$ will result in a circular trajectory on the surface of the Bloch sphere, giving a large error for most final times $\tf$. The trajectory will, however, periodically return to the north pole (orange-red curve in Fig. \ref{fig: trajectory illustration}). A general pulse (with nonconstant $\delta$), will typically give a nonperiodic trajectory that may not return to the north pole, resulting in a large population-transfer error outside of the adiabatic regime (dashed blue curve in Fig. \ref{fig: trajectory illustration}). 

The rest of this paper is organized as follows. In Sec.~\ref{sec: optimal pulse shape}, we project a multilevel system onto an effective two-level model. We then analyze errors due to noise and due to nonadiabatic transitions under a fast-QUAD pulse. In Sec.~\ref{sec: Canonical Models} we apply the analytical results from Sec.~\ref{sec: optimal pulse shape} to two generic models: the Landau-Zener model and a model with a constant energy gap (the ``constant-gap model''). In Sec.~\ref{sec: charge-qubit}, the results of Sec.~\ref{sec: optimal pulse shape} are applied to the experimentally relevant problem of readout for a double-quantum-dot charge qubit subjected to $1/f$ charge noise. Finally, in Sec.~\ref{sec: conclusions}, we present conclusions and future directions.

\section{Error model}
\label{sec: optimal pulse shape}

We consider a qubit initialized in one of the computational basis states $\ket{j}$ ($j=0,1$), see Fig.~\ref{fig: energy levels}. During a population transfer, there may be ``leakage'' into excited states, the qubit may become entangled with environmental degrees of freedom, and random noise sources may act on the system.  These effects lead (in general) to a mixed qubit density matrix at the final time $\tf$, $\rho_j(\tf)$. The population-transfer error $\epsilon_j$ associated with the initial state $\ket{j}$ is then
\begin{equation}
    \epsilon_j = 1-\mathrm{Tr}\left[\ket{j(\tf)}\bra{j(\tf)}\rho_j(\tf)\right],\,j=0,1.
    \label{eq: epsilon conversion error}
\end{equation}
The total error $\epsilon$ accounts for a prior probability (weight) $w_j$ for each initial state $\ket{j}$:
\begin{equation}\label{eq:error}
    \epsilon =\sum_{j=0,1} w_j\epsilon_j.
\end{equation}
In, e.g., the common case of equal priors, $w_j=1/2$ for $j=0,1$.

For a detailed analytic analysis, we now specialize to a two-level system, neglecting leakage errors. After qubit preparation, and before measurement, we take the qubit time evolution to be generated by a deterministic Hamiltonian $H_0(t)$ as well as a stochastic noise term caused by a random process $\boldsymbol{\eta}(t)=\left(\eta_x(t),\eta_y(t),\eta_z(t)\right)^T$:
\begin{eqnarray}
H_\eta(t) &=& H_0(t) + \frac{1}{2}\boldsymbol{\eta}(t)\cdot\boldsymbol{\sigma}, \label{eq: H_eta}\\
H_0(t) &=& \frac{1}{2} \boldsymbol{B}(t)\cdot\boldsymbol{\sigma}.\label{eq: H_0}
\end{eqnarray}
Here, $\boldsymbol{\sigma}=\left(\sigma_x,\sigma_y,\sigma_z\right)^T$ is the vector of Pauli operators. In what follows, we will restrict to the case where only two components of $\boldsymbol{B}(t)$ [$B_x(t),\,B_z(t)$] are time dependent to guarantee that the instantaneous eigenstates of $H_0(t)$ can be labeled by a single parameter.  Any constant (time-independent) nonzero component $B_y$ can be eliminated by an appropriate rotation, giving $\boldsymbol{B}(t)=\left(B_x(t),0,B_z(t)\right)^T$. The effective field $\boldsymbol{B}(t)$ can then be parametrized in terms of a magnitude and a phase: $B(t)e^{i\theta(t)}=B_z(t)+iB_x(t)$ and the instantaneous eigenstates can be written in terms of the single parameter $\theta(t)$: $\ket{j(t)}=\ket{j[\theta(t)]}$, allowing us to establish a fast-QUAD pulse [Eq.~\eqref{eq: diff eq}] for high-fidelity population transfer. The density operator $\rho_j(t_\mathrm{f})$ appearing in Eq.~\eqref{eq: epsilon conversion error} is  
\begin{equation}
    \rho_j(t_\mathrm{f}) = \llangle \mathcal{U}_\eta(t_\mathrm{f}) \ket{j}\bra{j} \mathcal{U}_\eta^\dagger(t_\mathrm{f}) \rrangle_\eta,
\end{equation}
where $\ket{j}$ is an eigenstate of $H_0(0)$, $\mathcal{U}_\eta(\tf)$ is a unitary operator [defined in Eq.~\eqref{eq:Ueta}, below] that is conditioned on the noise realization, and the average $\llangle\ldots\rrangle_\eta$ is performed over the random variables $\eta_\alpha(t),\,\alpha=x,y,z$. These variables are assumed to describe a stationary process with zero mean: $\llangle\eta_\alpha(t)\eta_\beta(t')\rrangle_\eta=\llangle\eta_\alpha(t-t')\eta_\beta(0)\rrangle_\eta$, $\llangle\eta_\alpha(t)\rrangle_\eta=0$. 

The noise-free time evolution of the qubit during a fast-QUAD pulse can be found exactly analytically (see Appendix \ref{sec: appendix generalized faquad}). The solutions are periodic circular trajectories of the Bloch vector on the surface of the Bloch sphere, see Fig.~\ref{fig: trajectory illustration}. In general, these trajectories may start and end at points that are misaligned with an instantaneous eigenstate (corresponding to the north and south poles of this Bloch sphere). To compensate for this misalignment, we additionally consider a preparation unitary $\mathcal{R}(0)$ before the pulse, and a pre-measurement unitary $\mathcal{R}(\tf)$ immediately following the pulse. The time-evolution operator $\mathcal{U}_\eta(\tf)$, including the preparation and pre-measurement unitaries is then
\begin{eqnarray}
    \mathcal{U}_\eta(\tf)&=&\mathcal{R}(\tf)U_\eta(\tf)\mathcal{R}^\dagger(0), \label{eq:Ueta}\\
    U_\eta(\tf)&=&\mathcal{T}e^{-i\int_0^{\tf}dt H_\eta(t)}.
\end{eqnarray}
Here, $\mathcal{T}$ is the usual time-ordering operator. The unitary $U_\eta(\tf)$ describes time evolution during the population transfer, conditioned on a single noise realization.

Our focus in this paper is on the case where the deterministic Hamiltonian $H_0(t)$ describes a fast-QUAD pulse [Eq.~\eqref{eq: diff eq}]. We consider two possibilities for the preparation/measurement unitary $\mathcal{R}(t)$:\\

\noindent
1. A generalized fast-QUAD protocol:
\begin{equation}\label{eq:generalizedFQ}
    \mathcal{R}(t)=R_y\left[\theta(t)\right]R_x^\dagger(\phi)R_y^\dagger\left[\theta(t)\right],
\end{equation}
2. The `standard' fast-QUAD protocol: 
\begin{equation} \label{eq:standardFQ}
    \mathcal{R}(t)=\mathbb{1}.
\end{equation}
Here, the angle $\phi=\arctan(\delta)$, shown in Fig.~\ref{fig: trajectory illustration}, is time-independent and $R_\alpha(\vartheta)=e^{-i\frac{\vartheta}{2}\sigma_\alpha}$ is a rotation. For the generalized protocol, this choice of $\mathcal{R}(t)$ rotates the initial state from an instantaneous eigenstate to an arc on the Bloch sphere and from the end of the arc back to an instantaneous eigenstate at an arbitrary final time. This leads to a vanishing noise-free error (see Appendix \ref{sec: appendix generalized faquad} for a derivation). It may not always be possible to implement a high-quality unitary $\mathcal{R}(t)$, in which case the standard protocol can still be used to reach a low population-transfer error under appropriate conditions (specifically, for special final times $\tf$ corresponding to full periods of the circular trajectory shown in Fig.~\ref{fig: trajectory illustration}).

\subsection{Population-transfer error}
In the generalized fast-QUAD protocol, noise-free nonadiabatic (Landau-Zener) transitions are counteracted by the application of the unitary $\mathcal{R}(t)$ [Eq.~(\ref{eq:generalizedFQ})]. The remaining population-transfer error is caused by the noise. To leading order in the noise, we find (see Appendix \ref{sec: appendix generalized faquad}):
\begin{eqnarray}
    \epsilon & \simeq & \epsilon_\eta(\delta),\\
    \epsilon_\eta(\delta) & = & \frac{1}{2} \sum_{\alpha,\beta}\int_{-\infty}^\infty \frac{d\omega}{2\pi}\frac{S_{\alpha\beta}(\omega)}{\omega^2}F_{\alpha\beta}(\omega,\tf),
    \label{eq:epsilon-eta-delta}
\end{eqnarray}
where $\alpha, \beta \in \{x,y,z\}$, and where the noise spectral density is
\begin{equation}
    S_{\alpha\beta}(\omega) = \int_{-\infty}^{\infty}dt\ e^{-i\omega t}\llangle\eta_\alpha(t)\eta_\beta\rrangle_\eta.
    \label{eq: S_alpha_beta noise spectrum}
\end{equation}
Here, we have introduced the generalized filter function $F_{\alpha\beta}(\omega,\tf)$ that depends on the fast-QUAD pulse through $\delta=\tan\phi$ and $\theta(t)$ (see Appendix \ref{sec: appendix generalized faquad} for a detailed definition). In the particular case of polarized noise, $\boldsymbol{\eta}(t)=[0,0,\eta_z(t)]$, there is only one nonvanishing term, $F_{zz}(\omega,\tf)=F(\omega,\tf)$:
\begin{align}
    &F(\omega,\tf) = \frac{\omega^2}{2}\left|  \int_0^{\tf}dt\ \xi(t) e^{-i\omega t} \right|^2,     \label{eq:filterfunctionpolarized}\\
    &\xi(t) = e^{-i\Phi_0(t)}\{\sin[\theta(t)] -i\sin(\phi)\cos[\theta(t)]\}.
\end{align}
The function $\Phi_0(t)$ is the dynamical phase acquired in the adiabatic frame
\begin{equation}\label{eq:Phi_0}
\Phi_0(t) = \int_0^t dt' \sqrt{B^2(t') + \dot{\theta}^2(t')}.
\end{equation}
Equation (\ref{eq:epsilon-eta-delta}) is valid in the limit of weak noise for any $\theta(t)$, even when no fast-QUAD protocol is used. Under the additional fast-QUAD constraint $\delta=\dot{\theta}/B$, and provided $\theta(t)$ is a monotonic function of time, the time integral in Eq.~\eqref{eq:Phi_0} can be traded for a parametric integral over $\theta$, giving:
\begin{equation}
    \Phi_0(t)  = \sqrt{\frac{\delta^2+1}{\delta^2}} \left|\Delta\theta(t)\right|, \label{eq:dynamicalphase}
\end{equation}
where $\Delta\theta(t)=\theta(t)-\theta(0)$. 

The form of Eq.~\eqref{eq:filterfunctionpolarized} is similar to that found in dynamical-decoupling theory \cite{martinis2003decoherence,beaudoin2015microscopic,cywinski2008enhance}, where here the term $\xi(t)$ replaces the ``sign function" $s(t)$  that is constrained to $s(t)=\pm 1$ under a sequence of $\pi$-pulses. As in the case of a dynamical-decoupling sequence, we can suppress contributions from the low-frequency part of the noise spectrum provided $\int_0^{\tf} dt\ \xi(t)=0$.  More generally, it is possible to minimize the error by minimizing the overlap of the filter function with the environmental noise spectrum, see Eq.~\eqref{eq:epsilon-eta-delta} (and  Fig.~\ref{fig: filter function illustration} for an example).

To assess the quality of a fast-QUAD population transfer in a number of numerical examples below, we consider a specific model of classical noise with amplitude $\sigma$ polarized along the $z$-direction, having a finite bandwidth. This model may arise from a quantum two-level systems (TLS) having a coherent precession frequency $\omega_0$ and an incoherent decay rate $\gamma$ \cite{shnirman2005low,galperin2006nonGaussian,schlor2019correlating}, leading to a Lorentzian spectral density $S_{\alpha\beta}(\omega) = S_{\omega_0}(\omega)\delta_{\alpha z}\delta_{\beta z}$:
\begin{equation}
    S_{\omega_0}(\omega) = \frac{\sigma^2 \gamma}{(\omega-\omega_0)^2+\gamma^2}+\frac{\sigma^2 \gamma}{(\omega+\omega_0)^2+\gamma^2}.
    \label{eq: OU Szz noise spectrum}
\end{equation}
In general, the TLS spectrum would also include a contribution at low frequency, centered at $\omega=0$, and the spectrum may generally be asymmetric as a function of $\omega$. The low-frequency contribution can be neglected if the qubit is only transversally coupled to the TLS [see Eq.~(7) in Ref.~\onlinecite{shnirman2005low}], and the spectrum can be taken to be symmetric about $\omega=0$ (as above) provided $\kB T$ is large compared to the TLS energy-level spacing, leading to approximately equal excitation and relaxation rates. A pure transversal coupling is natural for a qubit electrostatically coupled to TLSs having two states of differing electric dipole moment (e.g., ``left'' and ``right'' states of a symmetric double well) that are tunnel coupled, leading to bonding/antibonding TLS eigenstates.
For $\gg1$ independent TLSs coupled to a qubit, provided the noise variance is finite, the central-limit theorem guarantees that the noise will approximate a Gaussian process. In the limit of $\omega_0\to0$, this model then corresponds to a model of stationary Gaussian noise with a finite correlation time $\gamma^{-1}$ (an Ornstein-Uhlenbeck process \cite{uhlenbeck1930theory,wang1945theory}). For explicit numerical evaluation, we focus on the case $\omega_0=0$, but we also consider the case $\omega_0\ne 0$ in an example of noise mitigation, below.

In the standard fast-QUAD protocol, the population-transfer error may generally remain finite, even for vanishing noise, due to nonadiabatic (Landau-Zener) transitions. To estimate the error due to nonadiabatic transitions and due to noise, we perform a dual expansion in both the amplitude of the noise $\eta$ and in the adiabaticity parameter $\delta$, giving
\begin{equation}
    \epsilon = \tilde{\epsilon}(\delta)+\epsilon_\eta(0)+\mathcal{O}(\eta^2\delta,\eta^3).
\end{equation}
The contribution from the noise, $\epsilon_\eta(0)$, is given by the same expression [Eq.~\eqref{eq:epsilon-eta-delta}] as in the generalized protocol, but taken in the limit $\delta\to 0$. The contribution $\tilde{\epsilon}(\delta)$ gives the exact error in the limit of zero noise ($\eta\to 0$) and for any $\delta$ (see Appendix \ref{sec: appendix generalized faquad}):
\begin{equation}\label{eq:epsilon0}
    \tilde{\epsilon}(\delta) = \frac{\delta^2}{1+\delta^2}\sin^2\left[
    \sqrt{\frac{\delta^2+1}{\delta^{2}}} \frac{\dth(\tf)}{2}
    \right].
\end{equation}
When the dynamical phase acquired, Eq.~(\ref{eq:dynamicalphase}), is an integer multiple of $2\pi$ (corresponding to a complete traversal of the orange-red circle shown in Fig.~\ref{fig: trajectory illustration}), $\tilde{\epsilon}(\delta)=0$ identically, and the error is limited only by the noise term. We can exploit this behavior to design a population-transfer protocol with a final time $\tf$ that guarantees $\tilde{\epsilon}(\delta)=0$.

\section{Canonical models \label{sec: Canonical Models}}

\begin{figure}
\begin{centering}
\includegraphics[width=0.8\columnwidth]{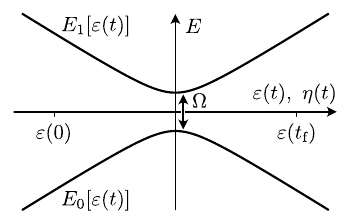}
\par\end{centering}
\caption{The Landau-Zener model is defined by a constant tunnel splitting $\Omega$ and a time-dependent detuning $\varepsilon(t)$, see Eq.~(\ref{eq:ModelParameterOmega}). }
\label{fig: landau zener levels}
\end{figure}

In this section, we illustrate the relevance of the fast-QUAD pulse [Eq.~(\ref{eq: diff eq})] for two ubiquitous canonical two-level models: (a) the Landau-Zener model, where the energy eigenvalues are described by hyperbolas as a function of a time-dependent detuning parameter $\varepsilon(t)$ (Fig.~\ref{fig: landau zener levels}), and (b) a ``constant-gap'' model, where the instantaneous adiabatic energy gap $B(t)=B$ is a constant, but where the eigenstates [determined by $\theta(t)$] are time dependent. Colored noise is included in the models, both analytically and numerically. We compare results for a fast-QUAD pulse to the results for a simple `linear' pulse. Finally, we compare the generalized fast-QUAD protocol (which relies on preparation and measurement unitaries) to the standard fast-QUAD protocol.

\subsection{Landau-Zener model \label{sec: Landau-Zener model}}

Both the Landau-Zener and constant-gap models are based on the two-dimensional Hamiltonian, Eq.~(\ref{eq: H_0}). For the Landau-Zener model, we refer to a detuning $\varepsilon(t)$ and to a tunnel splitting $\Omega$, which are related to the effective field components from Eq.~(\ref{eq: H_0}):  
\begin{eqnarray} \label{eq:ModelParameterEpsilon}
B_z(t) &=& \varepsilon(t),\\
B_x(t) &=& \Omega= \textrm{const}..\label{eq:ModelParameterOmega}
\end{eqnarray}

\begin{figure}
\begin{centering}
\includegraphics[width=1.0\columnwidth]{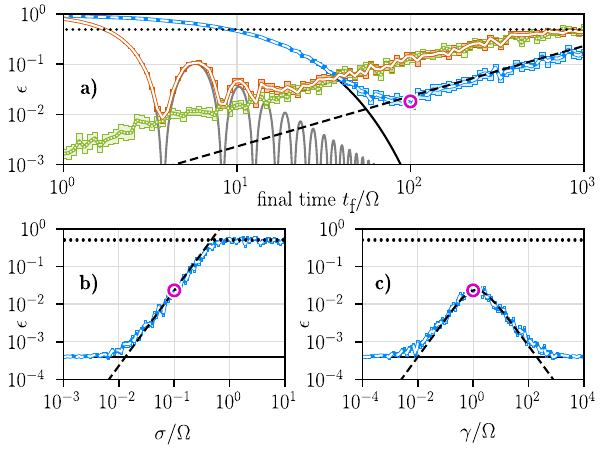}
\par\end{centering}
\caption{Numerically determined population-transfer error $\epsilon$ for the Landau-Zener model with a standard fast-QUAD protocol [without special preparation and measurement unitaries, Eq.~(\ref{eq:standardFQ})] (solid white line with an orange-red background), with a generalized fast-QUAD protocol [with special preparation and measurement unitaries, Eq.~\eqref{eq:generalizedFQ}] (dotted white line with a green background), and for a linear pulse (dashed white line with a blue background).
The shaded background indicates the error in the sample mean $\sigma_\epsilon/\sqrt{N_s}$ after $N_s=20$ random noise realizations, where $\sigma_\epsilon$ is the standard deviation of the population-transfer error. For these plots, we take $\varepsilon(\tf) = -\varepsilon(0) = 10\Omega$. We show the dependence on (a) the total sweep time $\tf$, (b) the noise amplitude $\sigma$, and (c) the switching rate $\gamma$ characterizing the Lorentzian noise, [Eq.~(\ref{eq: OU Szz noise spectrum}), with $\omega_0=0$]. The solid black lines give the usual error due to nonadiabatic transitions (in the absence of noise) for a linear infinite-time sweep [Eq.~(\ref{eq: LZ linear})].  The gray line gives the noise-free error for the fast-QUAD pulse [Eq.~(\ref{eq:epsilon0})]. The dashed black lines give the dominant error contribution in the adiabatic limit ($\Omega\tf\to\infty$) [Eq.~(\ref{eq:Kryzwda})], see Ref.~\cite{krzywda2020adiabatic} for a derivation. The value $\epsilon = 0.5$ (black dotted lines) corresponds to an equally probable outcome for $|0(\tf)\rangle$, $|1(\tf)\rangle$. The magenta circles correspond to the same set of parameters in each plot: $\sigma/\Omega = 0.1$, $\gamma/\Omega = 1$, $\tf\Omega = 100$.
The noise was simulated with an effective spectrum having a high-frequency cutoff satisfying $\omega_\tmax > 10 \max\{\varepsilon(\tf)/\hbar, \gamma\}$ (see Appendix \ref{sec: discretization of noise} for the precise choice of $\omega_\tmax$). In subfigure \textbf{c)}, we chose $\omega_\tmax > 10 \max\{\varepsilon(\tf), 10^3\Omega\}/\hbar$. 
}
\label{fig: landau zener model error}
\end{figure}
The tunnel splitting is taken to be constant, while the detuning parameter $\varepsilon(t)$ varies monotonically from an initial time $t=0$ to a final time $\tf$. 
We contrast two different cases, distinguished by the functional form of $\varepsilon(t)\in[\varepsilon(0),\varepsilon(\tf)]$: A linear pulse and a fast-QUAD pulse. The common linear pulse is given by:
\begin{equation}
\varepsilon_\textrm{linear}(t) =  \varepsilon(0) + [\varepsilon(\tf)-\varepsilon(0)] t/\tf. \label{eq:LZlinearPulse}
\end{equation}
Given the identifications for $\varepsilon$ and $\Omega$ in Eqs.~\eqref{eq:ModelParameterEpsilon} and \eqref{eq:ModelParameterOmega}, the fast-QUAD ansatz, Eq.~(\ref{eq: diff eq}), directly leads to a pulse shape
\begin{eqnarray}
\label{eq:fQpulse}
\varepsilon_\textrm{fQ}(t) &=& -\frac{(t+t_0)\Omega^2\delta}{\sqrt{1 - [(t+t_0)\Omega\delta]^2}},\\
t_0 &=& -\frac{1}{\Omega\delta}\frac{\varepsilon(0)}{\sqrt{\Omega^2+\varepsilon(0)^2}},\\
\delta &=& -\frac{1}{\Omega\tf}\left[
\frac{\varepsilon(\tf)}{\sqrt{\Omega^2+\varepsilon(\tf)^2}} - 
\frac{\varepsilon(0)}{\sqrt{\Omega^2+\varepsilon(0)^2}}
\right].\quad
\end{eqnarray}
In what follows, we analyze dynamics for a population transfer using the Landau-Zener model in two different contexts: In this section, we perform a symmetric sweep from $\varepsilon(0) = -10\Omega$ to $\varepsilon(\tf) = -\varepsilon(0) = 10\Omega$. In contrast, in Sec.~\ref{sec: charge-qubit} below, we consider a charge-qubit readout, where we pulse from the optimal operating point $\varepsilon(0)=0$ to a readout point where the two eigenstates are easily distinguishable, $\varepsilon(\tf) = 10\Omega$.

For the symmetric Landau-Zener sweep, numerical results for the population-transfer error are given in Fig.~\ref{fig: landau zener model error} for the linear pulse, Eq.~\eqref{eq:LZlinearPulse} (dashed white line with a blue background), and for the fast-QUAD pulses, Eq.~\eqref{eq:fQpulse}. The fast-QUAD result for the standard protocol is shown with a solid white line on an orange-red background and the result for the generalized protocol is shown with a dotted white line on a green background. These results are obtained from numerical integration of the time-dependent Schr\"odinger equation, accounting for detuning-noise realizations with the Lorentzian spectral density given by Eq.~(\ref{eq: OU Szz noise spectrum}), and assuming $\omega_0=0$; see Appendix~\ref{subsec: Noise} for details. We take the tunnel splitting $\Omega$ to be noise-free. The fast-QUAD pulse shows a significant advantage over the conventional linear pulse at short final times.

Analytic estimates of the population-transfer error (giving useful parametric dependencies) can be found in various limits. For example, the noise-free population-transfer error $\tilde{\epsilon}(\delta)$ for this (fast-QUAD) pulse is given by Eq.~(\ref{eq:epsilon0}) with $\Delta\theta(\tf)=2\arctan[\varepsilon(0)/\Omega]$. This contribution is indicated with a gray solid line in Fig.~\ref{fig: landau zener model error}. For the linear pulse, in the absence of noise, an exact formula for the population-transfer error is known for the finite-time Landau-Zener sweep \cite{vitanov1996landau}. However, for $|\varepsilon(0)|\gg |\Omega|$, the population-transfer error (ignoring noise) can be approximated more simply by the standard Landau-Zener formula \cite{landau1932theorie, zener1932non},
\begin{equation} \label{eq: LZ linear}
\epsilon_\textrm{LZ,linear} \simeq e^{-2\pi\Gamma};\quad 
\Gamma=\left(\frac{\Omega}{2}\right)^2 \frac{1}{|\dot{\varepsilon}|}=\frac{\Omega^2\tf}{8 |\varepsilon(0)|}.
\end{equation}
Equation \eqref{eq: LZ linear} is the dominant contribution to the total error for small $\tf$ and for weak noise. This contribution is given by solid black lines in Figs.~\ref{fig: landau zener model error}(a)-\ref{fig: landau zener model error}(c). In general, noise will introduce an additional contribution to the error. In the case of classical noise with a Lorentzian spectrum, Eq.~(\ref{eq: OU Szz noise spectrum}) with $\omega_0 = 0$, the noise contribution to the error has been found recently for the linear pulse and for $|\varepsilon(0)|\gg|\Omega|$ \cite{krzywda2020adiabatic}:
\begin{equation} \label{eq:Kryzwda} 
\epsilon_{\eta,\textrm{linear}} = \frac{\pi}{4}\frac{\Omega\sigma^2\tf}{\gamma|\varepsilon(0)|}
\left(
1 - \frac{1}{\sqrt{1 + \left( \frac{\gamma}{\Omega} \right)^2}}
\right).
\end{equation}
This source of error typically dominates for a long transfer time $\tf$ and is shown as black dashed lines in Figs.~\ref{fig: landau zener model error}(a)-\ref{fig: landau zener model error}(c). The minimum in the population-transfer error for a linear pulse as a function of sweep rate (equivalent to changing $\tf$ here) seen in Fig.~\ref{fig: landau zener model error}(a) (white dashed curve with a blue background) has also been discussed recently in Ref.~\onlinecite{krzywda2020adiabatic}. A similar nonmonotonic behavior of the nonadiabatic transition probability in the presence of a bosonic bath has also been found in, e.g., Ref.~\onlinecite{nalbach2009landau}.

We show the dependence of the population-transfer error on the two-level system rate $\gamma$ in Figure~\ref{fig: landau zener model error}(c). Here, we can observe two regimes: For small $\gamma/\Omega\ll 1$, $S_0(\omega)$ [see Eq.~(\ref{eq: OU Szz noise spectrum})] has very small weight at the minimal level splitting, $\omega\sim\Omega$. With increasing $\gamma$ (starting from $\gamma=0$), but still in the regime $\gamma/\Omega < 1$, $S_0(\omega\sim\Omega)\propto \gamma$ increases, leading to a corresponding increase in the error $\epsilon$.  On the other hand, for large $\gamma/\Omega > 1$, $S_0(\omega\sim\Omega)\propto 1/\gamma$ decreases, a consequence of  motional-averaging; in the motional-averaging regime, the noise averages out quickly on the time scale of evolution of the qubit. 

\subsection{Constant-gap model (Rabi model)\label{sec: Constant-gap model}}
\begin{figure}
\begin{centering}
\includegraphics[width=1.0\columnwidth]{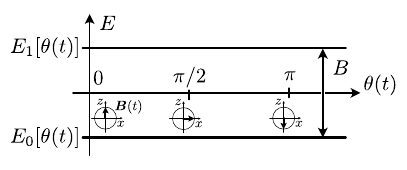}
\par\end{centering}
\caption{The constant-gap model is given by an effective field $\boldsymbol{B}(t)$ with constant magnitude $B=\textrm{const.}$ and a time-dependent angle $\theta(t) =  B\delta t =\Delta\theta(\tf) t/\tf$, see Eq.~(\ref{eq:CGB}). In simulations, we take $\Delta\theta(\tf) = \pi$.}
\label{fig: constant gap levels}
\end{figure}

\begin{figure}
\begin{centering}
\includegraphics[width=1\columnwidth]{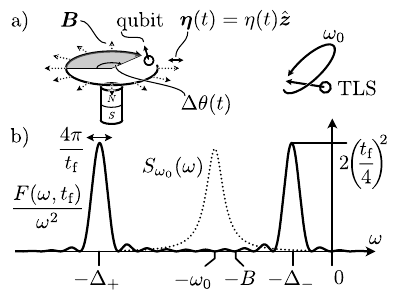}
\par\end{centering}
\caption{\label{fig: filter function illustration} \textbf{a)} Example realization of the constant-gap model: A qubit on a cyclic path in a magnetic field $\boldsymbol{B}$ pointing in the radial direction. In addition, we consider a source of noise due to a nearby two-level system (TLS) with precession frequency $\omega_0$. The dynamical phase acquired by the qubit is $\dth(t)$. \textbf{b)} Filter function $F(\omega, t_\textrm{f})$ (solid line) for the constant-gap model, see Eq.~(\ref{eq: CG filter function}), and noise spectrum $S_{\omega_0}(\omega)$ (dotted line) for a two-level system with central frequency $\omega=-\omega_0$ close to the qubit splitting, $B\sim\omega_0$. The two peaks of the filter function $F(\omega,\tf)$ are located at $\Delta_\pm = (\sqrt{1+\delta^{-2}}\pm1)B\delta$, see Eq.~(\ref{eq: CG filter function}). The final time $\tf$ is determined by $\tf = \dth(\tf)/B\delta$. The overlap between $F(\omega,\tf)$ and $S_{\omega_0}(\omega)$ can be minimized in the diabatic regime ($\delta \gg 1$) with $\dth(\tf)\gg\pi$, e.g. when the qubit revolves many times around the loop in subfigure \textbf{a)}.}
\end{figure}

\begin{figure}
\begin{centering}
\includegraphics[width=1.0\columnwidth]{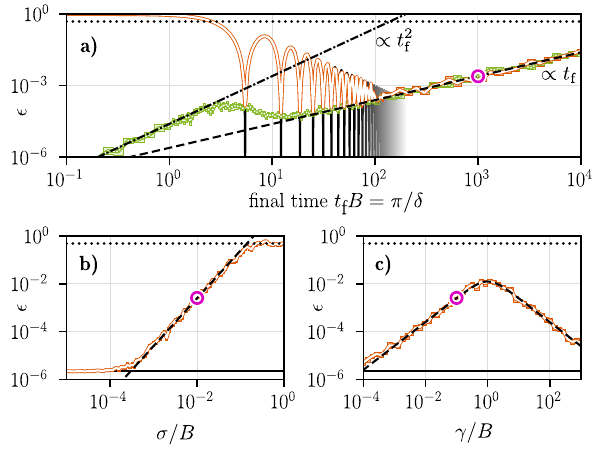}
\par\end{centering}
\caption{\label{fig: constant gap model}Population-transfer error $\epsilon$ for the constant-gap model [Eq.~(\ref{eq:CGdtheta})] with $\dth(\tf)=\pi$. Numerical results are shown for both the generalized protocol [Eq.~\eqref{eq:generalizedFQ}] (dotted line with green background) and for the standard protocol  [Eq.~\eqref{eq:standardFQ}] (solid line with orange-red background). The shaded background indicates the standard error in the sample mean after 20 random noise realizations. We show the dependence on \textbf{a)} the total time of the sweep $\tf$, \textbf{b)} the noise amplitude $\sigma$, and \textbf{c)} the switching rate $\gamma$ [Eq.~\eqref{eq: OU Szz noise spectrum} with $\omega_0 = 0$]. In each subfigure, the solid black line shows $\tilde{\epsilon}(\delta)$, Eq.~(\ref{eq:epsilon0}), the dashed black line shows the adiabatic limit ($\delta \to 0$), Eq.~(\ref{eq: CG error adiabatic limit}), while the dash-dotted line gives the diabatic limit ($\delta\to\infty$) for the generalized protocol, Eq.~(\ref{eq: CG error diabatic limit}). The value $\epsilon=0.5$ (dotted lines) corresponds to an equally probable outcome for $\ket{0(\tf)}, \ket{1(\tf)}$. The magenta circles correspond to the same set of parameters in each plot: $\sigma/B = 0.01$, $\gamma/B = 0.1$, $\tf B = 10^3$. For these simulations, the high-frequency cutoff was chosen such that $\omega_\tmax > 10\max\{B/\hbar, \gamma\}$ (see Appendix \ref{sec: discretization of noise} for the precise choice). In subfigure \textbf{c)}, we chose $\omega_\tmax$ satisfying $\omega_\tmax > 10^4 B/\hbar$. }
\label{fig: constant gap model error}
\end{figure}

In this section, we consider the ``constant-gap model'' (Fig.~\ref{fig: constant gap levels}), where the qubit splitting (magnitude of the effective magnetic field), $B$, is a constant, but where the direction of the effective magnetic field may vary in the $x$-$z$ plane: $\boldsymbol{B}(t)=B\left[\sin\theta(t),0,\cos\theta(t)\right]$. The time-dependent Hamiltonian $H_0(t)=\frac{1}{2}\boldsymbol{B}(t)\cdot\boldsymbol{\sigma}$ for this case arises naturally, for example, from the Rabi Hamiltonian for a driven qubit with a constant drive amplitude $\propto B$, but with a varying phase $\theta(t)$, $H_R(t)=\frac{1}{2}\omega_q\sigma_y+2 B\cos\left[\omega t+\theta(t)\right]\sigma_z$. In a rotating frame and at zero detuning ($\omega=\omega_q$), and within a rotating-wave approximation (for $\omega_q\gg B$), we have $U H_R U^\dagger -i U \dot{U}^\dagger \simeq H_0(t)$, with $U=e^{i\frac{1}{2}\sigma_y \omega_q t}$. A schematic illustration of an alternative physical setup for this model is shown in Fig.~\ref{fig: filter function illustration}(a). This model may also apply to shuttling of electron or hole spins through a region with a spatially varying g-tensor, as realized recently for electron-spin qubits in silicon \cite{yoneda2021coherent}, or through a region with a spatially varying effective hyperfine field, as realized for spin qubits in GaAs \cite{mortemousque2021enhanced}. Finally, this model may also be relevant for shuttling of ion-trap qubits through spatially varying and noisy electric and magnetic fields \cite{kaushal2020shuttling,pino2021demonstration}.

We now further restrict to a fast-QUAD pulse, for which the polar angular velocity $\dot{\theta}= B\delta$ is constant:
\begin{eqnarray}
\dot{\theta} &=& B\delta = \frac{\dth(\tf)}{\tf} = \textrm{const.},\label{eq:CGdtheta}\\
\boldsymbol{B}(t) &=& B\ (\sin(B\delta t), 0, \cos(B\delta t))^T.\label{eq:CGB}
\end{eqnarray}
In the adiabatic regime ($\delta\ll 1$), the qubit pseudospin-1/2 will precess many times in the effective field $\boldsymbol{B}(t)$ as it executes a cycle. In the diabatic regime ($\delta \gg 1$), the effective magnetic field will execute many cycles on the timescale of qubit precession. 

Due to the simplicity of the effective-magnetic-field evolution under a fast-QUAD pulse, the constant-gap model is an ideal testbed for a comparison of the standard [Eq.~(\ref{eq:standardFQ})] and generalized [Eq.~(\ref{eq:generalizedFQ})] fast-QUAD protocols. These two protocols are compared directly in Fig.~\ref{fig: constant gap model error}(a): The population-transfer error for the standard protocol (solid white line on an orange-red background) and the error for the generalized protocol (dotted white line on a green background) have been found numerically. The noise-free error contribution $\tilde{\epsilon}(\delta)$ [solid black lines in Figs.~\ref{fig: constant gap model error}(a)-\ref{fig: constant gap model error}(c)], Eq.~(\ref{eq:epsilon0}), is completely eliminated in the generalized protocol. The error contribution arising from noise $\epsilon_\eta(\delta)$, Eq.~(\ref{eq:epsilon-eta-delta}), enters for both the generalized and standard protocols.
Here, we have assumed a source of polarized noise along the $z$-direction $\boldsymbol{\eta}(t) = (0,0,\eta(t))^T$. With this assumption, we analytically calculate the exact filter function $F(\omega,\tf)$ given in Eq.~(\ref{eq:filterfunctionpolarized}) for the constant-gap model [see Appendix~\ref{app: filterfunction} resulting in Eq.~(\ref{eq: CG filter function})]. 

The filter function can be analyzed in two simple limits: the adiabatic limit ($\delta\rightarrow 0$) and the diabatic limit ($\delta\rightarrow\infty$). For fixed $\dth(\tf)$, in the adiabatic limit ($\delta\to 0$), $F(\omega,\tf)/\omega^2$ becomes strongly peaked about $\omega=-B$ with full width $\propto 1/\tf\propto 4\pi B\delta/\dth(\tf)\to 0$ and height $\propto t_f^2 \propto \left[\dth(\tf)/4B\delta\right]^2\to\infty$ [see Fig.~\ref{fig: filter function illustration}(b)], allowing us to replace the filter function asymptotically with a delta function having the same weight as $F(\omega,\tf)/\omega^2$. Using the relationship between $\tf$ and $\delta$ given in Eq.~\eqref{eq:CGdtheta}, we find:
\begin{eqnarray}
\frac{F(\omega,\tf)}{\omega^2} &\sim & \frac{\pi}{2B\delta}W[\dth(\tf)]\delta(\omega+B)\quad (\delta\to 0),\label{eq: CG FF adiabaic limit}\\
W[\dth(\tf)]&=&\dth(\tf) - \frac{1}{2}\sin[2\dth(\tf)].
\end{eqnarray}
We use the symbol ``$\sim$'' here to indicate an asymptotic equality. Inserting equation \eqref{eq: CG FF adiabaic limit} into the expression for the transfer error in terms of the noise overlap [Eq.~(\ref{eq:epsilon-eta-delta})], using the relation given in  Eq.~\eqref{eq:CGdtheta}, and choosing the boundary condition $\dth(\tf)=\pi$, gives
\begin{equation}
\epsilon_\eta(\delta\to 0) \sim \frac{\pi}{8B\delta} S_{zz}(-B) = \frac{\tf}{8} S_{zz}(-B).
\label{eq: CG error adiabatic limit}
\end{equation}
For explicit numerical and analytical evaluation of the state-transfer error shown in Fig.~\ref{fig: constant gap model error}, we consider $S_{zz}(\omega)=S_{\omega_0=0}(\omega)$, with $S_0(\omega)$ given by Eq.~\eqref{eq: OU Szz noise spectrum} (a Lorentzian spectrum centered at zero frequency). In the diabatic limit [$\delta\to\infty$ at fixed $\dth(\tf)$], we find the leading behavior of the filter function: 
\begin{equation}
\frac{F(\omega,\tf=\dth(\tf)/B\delta)}{\omega^{2}} = \frac{1}{2}\left(\frac{\dth(\tf)}{B\delta}\right)^2 + \mathcal{O} \left( \frac{\omega^2\pi^4}{B^4\delta^4} \right).
\label{eq:F_small_tf}
\end{equation}
Provided the noise spectral density has a finite integrated weight, the conversion error due to the noise, Eq.~(\ref{eq:epsilon-eta-delta}), can be approximated by the leading behavior for $\delta\to\infty$ (equivalently, $\tf\to 0$) after substituting Eq.~\eqref{eq:F_small_tf}:
\begin{equation}
\epsilon_\eta\left(\delta=\frac{\dth(\tf)}{B\tf}\to\infty\right) \sim \left(\frac{\tf}{2}\right)^2 \int_{-\infty}^\infty \frac{d\omega}{2\pi}S_{zz}(\omega).
\label{eq: CG error diabatic limit}
\end{equation}
As expected, the analytic error contributions [Eqs.~\eqref{eq:epsilon0}, \eqref{eq: CG error adiabatic limit}, and \eqref{eq: CG error diabatic limit}] reproduce the numerical evaluation in appropriate limits (see Fig.~\ref{fig: constant gap model}).

In the adiabatic limit, the error is dominated by the contribution to the noise at the qubit splitting, $\epsilon\propto S_{zz}(\omega = -B)$. For example, this proportionality describes both the $\sigma$-dependent error shown in Fig.~\ref{fig: constant gap model}(b) and the $\gamma$-dependent error in Fig.~\ref{fig: constant gap model}(c). Thus, if the noise spectrum could be properly manipulated, it would be possible to reduce the error. Alternatively, it is often possible to modify the population-transfer protocol to adjust the filter function $F(\omega,\tf)$ according to the noise spectrum, as in the example that follows.

An important case to consider is an environmental quantum TLS producing noise with a spectrum $S_{\omega_0}(\omega)$ [Eq.~\eqref{eq: OU Szz noise spectrum}] centered around a central frequency $\omega_0$. Although there are generally many TLSs in the environment, if one TLS has a frequency $\omega_0$ close to the qubit splitting, $\omega_0\simeq B$, see Fig.~\ref{fig: filter function illustration}, then this particular TLS may provide the dominant contribution to the error for an adiabatic protocol ($\tf\to\infty$), where the filter function is flat as a function of $\omega$. Outside of the adiabatic regime, the filter function $F(\omega,\tf)$ of the constant-gap model, Eq.~(\ref{eq: CG filter function}), is instead peaked around the frequencies $\omega=-\Delta_\pm$ with
\begin{equation}
\Delta_\pm = \left(\sqrt{1+\delta^{-2}}\pm 1\right)B\delta.
\end{equation}
The peaks have a full-width $ 4\pi/\tf=4\pi B\delta/\dth(\tf)$. By reducing the overlap between the noise spectrum $S_{\omega_0}(\omega)$ and the filter function $F(\omega,\tf)/\omega^2$, the population-transfer error $\epsilon_\eta(\delta)$ [Eq.~(\ref{eq:epsilon-eta-delta})] can be improved. For the overlap to be small, we require $|\omega_0-\Delta_\pm|\gg \mathrm{max}\{2\pi/\tf,\gamma\}$. For $2\pi/\tf>\gamma$ and $\omega_0\sim B$ in the diabatic regime ($\delta\gg 1$), this condition implies $\Delta\theta(\tf)\gg \pi$, i.e., we require that the two-level system precesses many times during the sweep. This leads to fast averaging of the noise, in analogy with a repeated dynamical decoupling sequence.

\section{Charge-qubit readout\label{sec: charge-qubit}}

\begin{figure}
\begin{centering}
\includegraphics[width=1.0\columnwidth]{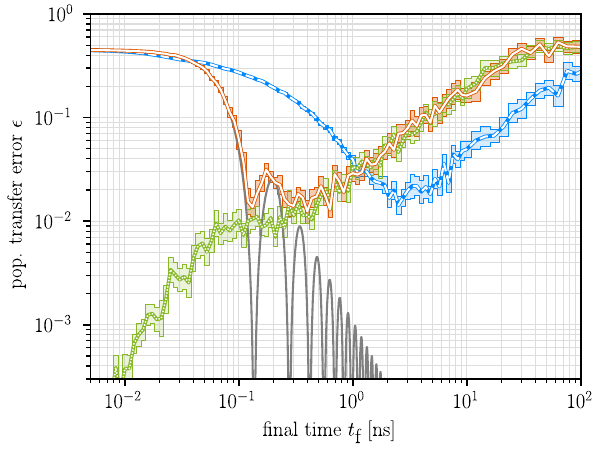}
\par\end{centering}
\caption{Population transfer error $\epsilon$ for the readout of a charge qubit with detuning $\varepsilon(t)$ and tunnel splitting $\Omega$. The qubit is operated at an optimal point, $\varepsilon(0)\simeq 0$ and measured at $\varepsilon(\tf)>\Omega$. The population-transfer error is calculated for a linear pulse (white dashed line with blue background), the standard fast-QUAD protocol (white line with orange-red background), and the generalized protocol (white dotted line with green background). The shaded background indicates the error in the sample mean after $N_s=20$ random noise realizations. The solid gray line shows the noise-free error for the standard fast-QUAD pulse [Eq.~\eqref{eq:epsilon0}]. The detuning at the measurement point is $\varepsilon(\tf) = 200~\ueV$ and the tunnel splitting is $\Omega = 20~\ueV$. The noise $\eta(t)$ is given by a $1/f$ spectrum [Eq.~\eqref{eq: 1_f spectrum}], where the parameter $A=2~\ueV^2$ is consistent with both experimentally measured charge-qubit dephasing times and with charge noise extracted from transport measurements on nanostructures (see text). In practice, for simulations we incorporate the noise spectral weight in the range from $\omega/2\pi=\omegaLow/2\pi = 1~\Hz$ to $\omega/2\pi=\omega_\tmin/2\pi = 1~\MHz$ into a quasistatic term $S(\omega)\sim \delta(\omega)$. We simulate the remaining dynamic [$S(\omega)\sim 1/\omega$] contributions for $\omega_\tmin <\omega < \omega_\tmax$ [with $\omega_\tmax>10\varepsilon(\tf)/\hbar$]. See Appendix \ref{sec: discretization of noise} for details.}
\label{fig: charge qubit}
\end{figure}

In this section, we simulate the population-transfer process for a charge qubit defined in the two lowest-energy orbitals of a single electron in a double quantum dot. The energy levels for this problem map directly onto the Landau-Zener model studied in Sec.~\ref{sec: Landau-Zener model} with a detuning parameter $\varepsilon$ controlling the asymmetry of the double-dot potential and a tunnel splitting $\Omega$ controlled by the overlap of single-particle orbitals in the two dots. In these simulations, we further account for $1/f$ charge noise with parameters that are consistent with recent experiments, illustrating the directly realizable advantages of fast-QUAD protocols in a realistic experimental system (see Fig.~\ref{fig: charge qubit}). We expect the advantages illustrated here to translate directly to analogous systems (e.g., spin-to-charge conversion for singlet-triplet qubits defined in two-electron spin states or parity-to-charge conversion schemes for Majorana qubits).

Although double-quantum-dot charge qubits \citep{hayashi2003coherent, petersson2010quantum} typically show short coherence and relaxation times relative to spin qubits, they also serve as an important platform for hybrid spin-charge qubits \cite{shi2012fast,kim2014quantum} that allow for rapid electrical control. The large transition dipole afforded by these systems also allows for strong-coupling effects when coupling to microwave cavities, while unwanted noise sensitivity can be minimized under carefully chosen operating conditions \citep{scarlino2021situ}. These qubits are best operated at a noise-insensitive point corresponding to a symmetric double-dot potential [$\varepsilon(0) = 0$ in the Landau-Zener model], where the two energy eigenstates are least distinguishable based on the charge distribution. For single-shot readout, it is advantageous to convert from the symmetric configuration to an asymmetric double-dot potential [$\varepsilon(\tf)\gg\Omega$], where the two energy eigenstates are distinguished by an additional electron charge on one quantum dot or the other. These two states can be more easily differentiated with a proximal charge sensor (see, e.g., Ref.~\onlinecite{petersson2010quantum}).

In simulations (Fig.~\ref{fig: charge qubit}), we choose the following values for the tunnel splitting and the detuning at the measurement point:
\begin{eqnarray}
\Omega &=& 20~\ueV,\\
\varepsilon(\tf) &=& 200~\ueV.
\end{eqnarray}
The value chosen for $\Omega$ is roughly consistent with the tunnel coupling $\Omega/2=8~\ueV$, reported in Ref.~\onlinecite{petersson2010quantum}. The detuning at the measurement point was chosen to satisfy $\varepsilon(\tf) = 10\Omega$, guaranteeing that $\varepsilon\gg \Omega$ at the measurement point so that the two energy eigenstates at this point are maximally distinguishable in charge.

In a wide range of devices, charge noise is commonly measured to show a $1/f$ spectrum over a broad frequency range \citep{yoneda2018aQuantum, petit2018spin}, which is expected to arise from an ensemble of two-level fluctuators in the proximity of a charge qubit \citep{dutta1981low,schriefl2006decoherence,paladino20141}. Rather than simulating the noise due to a single two-level system as in the previous sections, here we simulate $1/f$-noise: 
\begin{equation}
S(\omega) = \begin{cases}
\frac{A}{|\omega|} , &|\omega|>\omegaLow, \\
0, &\textrm{otherwise.}
\end{cases} \label{eq: 1_f spectrum}
\end{equation}
There are well-known complications in the $1/f$ noise spectrum at low and high frequencies. Any measurement of $S(\omega)$ performed over a finite measurement time $T_\mathrm{m}$ only has access to Fourier components $\omega\lesssim \omegaLow = 2\pi/T_\mathrm{m}$, so we have introduced a low-frequency cutoff in Eq.~\eqref{eq: 1_f spectrum}. In simulations, we take $\omegaLow/2\pi=1\,\mathrm{Hz}$, consistent with $T_\mathrm{m}=1\,\mathrm{s}$. In practice, $T_\mathrm{m}$ is the time scale for recalibration of the detuning $\varepsilon$, which is otherwise affected by random fluctuations $\varepsilon\to\varepsilon+\eta(t)$. The integrated spectral weight under $S(\omega)$ gives the variance in the noise parameter $\eta$. If this variance is to be finite, the true noise spectrum must also roll off faster than $1/\omega$ above some characteristic frequency, set by the shortest time scale for two-level fluctuator jumps. However, as we show below, provided the fastest fluctuator time scales are short compared to the self-consistently determined dephasing time $\tt$, the particular value of the high-frequency cutoff becomes irrelevant so we do not include it in the model.

We establish a realistic numerical value of the parameter $A$ in Eq.~\eqref{eq: 1_f spectrum} by relating this parameter to the experimentally measured dephasing time $\tt$ found for charge qubits. We then confirm that the value of $A$ extracted from $\tt$ is consistent with independent measurements of this parameter based on voltage fluctuations in nanoscale devices.

For a double-dot charge qubit, the dephasing time $\tt$ far from the optimal operating point ($\varepsilon>\Omega$) can be found from a Ramsey sequence. First, the qubit is prepared in the molecular ground state at $\varepsilon=0$, then after a rapid pulse to $\varepsilon>\Omega$, the qubit evolves freely for a time $t$. Another rapid pulse then returns the detuning to $\varepsilon=0$, after which an adiabatic pulse from $\varepsilon=0$ to $\varepsilon\gg \Omega$ can be used to read out the qubit. For this protocol, during the free evolution time the qubit coherence acquires a random phase, $\phi_\eta(t) = \int_0^t dt'\eta(t')$. When $\eta(t)$ is a stationary Gaussian random variable described by the spectrum $S(\omega)$, then the Ramsey sequence described above measures decay in the off-diagonal element of the qubit density matrix, proportional to the coherence factor
\begin{equation}
    C(t)=\left<e^{i\phi_\eta(t)}\right>_\eta=e^{-\half\xct{\phi^2(t)}_\eta},
\end{equation}
with a phase variance given by
\begin{equation}
\xct{\phi^2(t)}_\eta = \int_{-\infty}^\infty \frac{d\omega}{2\pi}\frac{\sin^2(\omega t/2)}{(\omega/2)^2} S(\omega). \label{eq: ramsey}
\end{equation}
After inserting the $1/f$ noise spectrum [Eq.~\eqref{eq: 1_f A}], we find the following asymptotic form for the integral, Eq.~\eqref{eq: ramsey}:
\begin{equation}\label{eq:AsymptoticPhiVariance}
    \xct{\phi^2(t)}_\eta\sim \frac{A t^2}{\pi}\ln\left(\frac{1}{\omegaLow t}\right);\quad \omegaLow t\to 0.
\end{equation}
The corrections are logarithmic and so the formula is only accurate when $\ln(1/\omegaLow t)\gg 1$.

We define the dephasing time $\tt$ to be the time at which $C(t)$ is suppressed to $e^{-1}$:
\begin{equation}\label{eq:T2StarDef}
    \half\xct{\phi^2(\tt)}_\eta = 1.
\end{equation}
Provided $\ln(1/\omegaLow \tt)\gg 1$, we can substitute the asymptotic expression given in Eq.~\eqref{eq:AsymptoticPhiVariance} into Eq.~\eqref{eq:T2StarDef} to solve for $A$ in terms of the measured quantity, $\tt$ (we also restore $\hbar$):
\begin{equation}
    A \simeq \frac{2\pi\hbar^2}{(\tt)^2\ln\round{\frac{1}{\omegaLow\tt}}}. \label{eq: A eq}
\end{equation}
With the same (logarithmic) accuracy, Eq.~\eqref{eq: A eq} can be inverted to give an expression for $\tt$ in terms of $A$ and $\omegaLow$ \citep{cottet2001superconducting,makhlin2004dissipative,yang2019achieving} (see, e.g., Eq.~(35) of Ref.~\onlinecite{makhlin2004dissipative}).

The Ramsey experiment described above has been performed in Ref.~\onlinecite{petersson2010quantum} for a double-quantum-dot charge qubit containing a single electron. Both the Ramsey measurements (with reported measurement time $T_\mathrm{m}=100\,\mathrm{ms}$, corresponding to $\omegaLow/2\pi=10\,\mathrm{Hz}$) and photo-assisted tunneling linewidth measurements in the same work are consistent with $\tt\simeq 250\,\mathrm{ps}$. Inserting these values for $\tt$ and $\omegaLow$ into Eq.~\eqref{eq: A eq} gives an estimate for the noise amplitude in these experiments:
\begin{equation}
A \approx 2~\ueV^2. \label{eq: 1_f A}
\end{equation}
To confirm that this value is typical, we compare to the typically cited quantity $\sqrt{S_0} = \sqrt{S(\omega=2\pi\times 1~\Hz)}$:
\begin{equation}
    \sqrt{S_0} = \sqrt{\frac{A}{2\pi}\Hz\inv} \approx 0.6~\ueV/\sqrt{\Hz}. \label{eq: S_0 derived}
\end{equation}
This lies in the range of values ($\sqrt{S_0} \simeq 0.3-2~\ueV/\sqrt{\Hz}$) reported for different devices \cite{freeman2016comparison,struck2020low,kranz2020exploiting}. For example, the authors of Ref.~\cite{freeman2016comparison} have found a range of values, from $\sqrt{S_0}\simeq 0.49\, \ueV/\sqrt{\Hz}$ to $\sqrt{S_0}\simeq 2.1\, \ueV/\sqrt{\Hz}$ from current fluctuations through single-electron transistors (SETs) defined in various devices based on Si/SiO$_2$ and Si/SiGe heterostructures. The authors of Ref.~\onlinecite{struck2020low} report $\sqrt{S_0}=0.47\,\ueV/\sqrt{\Hz}$ from SET current fluctuations in a Si/SiGe device (see the caption of Fig.~3, Ref.~\onlinecite{struck2020low}). An overview of reported measurements for $S_0$ in a broader range of devices is given in Ref.~\onlinecite{kranz2020exploiting}, Table 1.

Figure \ref{fig: charge qubit} shows the population-transfer error $\epsilon(\tf)$ for a charge qubit. A time-linear pulse results in a similar population-transfer error (white-dashed line with blue background) as in the Landau-Zener model. The standard fast-QUAD protocol (white line with orange-red background in Fig.~\ref{fig: charge qubit}) leads to minima in the transfer error at specific final times [Eq.~\eqref{eq:epsilon0}]. The generalized protocol (white dotted line with green background) removes the contribution to the population-transfer error arising from nonadiabatic transitions, as we have also seen for the Landau-Zener model, Fig.~\ref{fig: landau zener model error}. Figure \ref{fig: charge qubit} indicates that the fast-QUAD protocol could lead to a reduction in the readout pulse time by an order of magnitude, with no reduction in the state conversion error, relative to the linear pulse. Alternatively, when fast preparation and measurement unitaries are available, the generalized fast-QUAD protocol could lead to a significant reduction in the population transfer error, leading to a higher quality readout.

\section{Conclusions\label{sec: conclusions}}

In this paper, we have introduced a formalism that allows us to simultaneously minimize nonadiabatic errors while mitigating noise sources. In particular, we have derived closed-form analytic expressions for the error under a fast-QUAD protocol accounting for a generalized filter function that includes the influence of the shaped pulse as well as anisotropic classical noise. Further, we have introduced a generalized protocol that can achieve zero error in the absence of noise for a two-level system, provided additional high-quality initialization and measurement unitaries are available. Moreover, we have performed a detailed analysis of the effects of noise on population transfer during a fast-QUAD pulse. This numerical analysis allowed us to demonstrate the utility of the filter-function formalism, analogous to that regularly employed in dynamical-decoupling theory. The filter-function formalism provides a natural framework for designing pulses that avoid both the detrimental effects of noise and of unwanted nonadiabatic transitions. We have applied these ideas first to two generic and widely used models (the Landau-Zener model and the constant-gap model) and then to a realistic charge-qubit readout. 

The analysis presented here can be directly applied to a wide range of quantum systems, where the goal is to transfer population from one eigenbasis to another eigenbasis that is related by a single parameter $\theta(t)$. To demonstrate such an application, we simulated the readout of a charge qubit accounting for noise with a realistic ($1/f$) spectrum and noise amplitude that is typical of current experimental devices. For the charge-qubit example, we find that the fast-QUAD pulse can be used to significantly reduce the pulse time  compared to a linear pulse, while maintaining a comparable population-transfer error. When high-quality preparation and measurement unitaries are available, the generalized fast-QUAD pulse can significantly reduce both the pulse time and readout error in this context. Further examples that can benefit from fast-QUAD pulses include spin-to-charge conversion schemes, shuttling of spin qubits in quantum dots or of ion-trap qubits, and parity-to-charge conversion schemes for Majorana zero modes.

Extensions of this work could incorporate fast-QUAD pulses into quantum gates or a quantum memory, making use of the strategies we have presented to simultaneously suppress noise, nonadiabatic transitions, and leakage errors. Another possible extension would be to investigate the population-transfer error under the influence of quantum noise while incorporating the fast-QUAD protocol. Some work has been done to incorporate a quantum environment into Landau-Zener dynamics using more standard pulse shapes \cite{nalbach2009landau,krzywda2021interplay}, but there may be subtleties arising from the specific pulse shapes taken here and their ability to cancel only classical-noise features while potentially leaving quantum contributions. For example, a Carr-Purcell dynamical decoupling sequence generally leads to no acquired phase for a qubit undergoing pure dephasing due to a classical environment, but a finite phase can arise under such a sequence for a quantum environment associated with a noncommuting bath operator \cite{paz2017multiqubit,kwiatkowski2020influence,wang2021intrinsic,mcintyre2022non}. Another natural extension of this work would be to use the filter-function formalism derived here to design a new class of pulses that fully accounts for the noise spectrum, allowing for a simultaneous minimization of both noise-induced errors and nonadiabatic errors. 

\begin{acknowledgments}
We acknowledge support from the Natural Sciences and Engineering Research Council of Canada (NSERC), the Fonds de recherche du Québec -- Nature et technologies (FRQNT), and the National Research Council of Canada's Quantum Sensors Challenge Program (QSP). This research was undertaken thanks in part to funding from the Canada First Research Excellence Fund.
\end{acknowledgments}

\appendix

\section{Population-transfer error\label{sec: appendix generalized faquad}}
In this Appendix, we derive analytical expressions for the population-transfer error [Eqs.~(\ref{eq: epsilon conversion error}) and (\ref{eq:error}) of the main text] for two protocols: (1)~the generalized fast-QUAD protocol that incorporates additional preparation and measurement unitaries, and (2)~the standard fast-QUAD protocol, where preparation/measurement unitaries are excluded.

We first diagonalize the noise-free time-evolution operator with a unitary $S(t)$, leading to an expression for the noisy evolution:
\begin{equation}\label{eq:UetaAppendix}
    U_\eta(t)=S(t)e^{-\frac{i}{2}\Phi_\eta(t)\sigma_z}e^{-i\int_0^t dt' V_\eta (t')}S^\dagger(0),
\end{equation}
where $S(t)$ is an SU(2) rotation that can be written in terms of two Euler angles, $\theta(t)$ and $\phi$:
\begin{equation}
    S(t) = R_y[\theta(t)]R_x(\phi);\quad R_\alpha(\vartheta)=e^{-\frac{i}{2}\vartheta\sigma_\alpha}, 
\end{equation}
and where
\begin{equation}
    \phi = \arctan{\delta}.
\end{equation}
The phase $\Phi_\eta(t)$ is given by
\begin{equation}
    \Phi_\eta(t) =  \Phi_0(t)+\delta\Phi_\eta(t),
\end{equation}
with
\begin{eqnarray}
    \Phi_0(t) & = & \int_0^t dt'\sqrt{B^2(t')+\dot{\theta}^2(t')}\\
    & = & \sqrt{1+\delta^{-2}}\left|\Delta\theta(t)\right|,
\end{eqnarray}
where $\Delta\theta(t)=\theta(t)-\theta(0)$, and in the second line above we have used the fast-QUAD relationship $\delta = \dot\theta(t)/B(t)$ [Eq.~\eqref{eq: diff eq}], and the additional condition that $\theta(t)$ is a monotonic function of time $t$. The noise enters through the phase in
\begin{equation}
    \delta \Phi_\eta(t) = \int_0^{t} dt \tilde{\eta}_z(t),
\end{equation}
and through the perturbation $V_\eta(t)$:
\begin{equation}
    V_\eta(t) = \frac{1}{2}\left(\tilde{\eta}_+(t)e^{-i\Phi_\eta(t)}\sigma_-+\tilde{\eta}_-(t)e^{i\Phi_\eta(t)}\sigma_+\right),
\end{equation}
with $\tilde{\eta}_\pm(t) = \tilde{\eta}_x(t)\pm i\tilde{\eta}_y(t)$. Finally, the parameters $\tilde{\eta}_\alpha(t)$ are related to the original noise variables $\eta_\alpha(t)$, through a rotation:
\begin{equation}\label{eq:etaRotation}
    \tilde{\boldsymbol{\eta}}(t)\cdot\boldsymbol{\sigma} = S^\dagger(t)\boldsymbol{\eta}(t)\cdot\boldsymbol{\sigma} S(t).
\end{equation}
In particular, Eq.~\eqref{eq:etaRotation} implies $\tilde{\boldsymbol{\eta}}(t)=\left[M(t)\right]\cdot\boldsymbol{\eta}(t)$ where
\begin{equation}
    \left[M(t)\right] = \begin{pmatrix}
\cos\theta(t) & 0 & -\sin\theta(t)\\
\sin\phi\sin\theta(t) & \cos\phi & \sin\phi\cos\theta(t)\\
\cos\phi\sin\theta(t) & -\sin\phi & \cos\phi\cos\theta(t)
\end{pmatrix}.
\label{eq:Mmatrix}
\end{equation}

\subsection{Generalized fast-QUAD protocol}
For the generalized fast-QUAD protocol, we minimize the error for $|\boldsymbol{\eta}|=\eta\to 0$ with the choice:
\begin{eqnarray}
    \mathcal{R}(t) &=& R_y\left[\theta(t)\right]S^\dagger(t)\label{eq:Rt-in-terms-of-S}\\
    &=& R_y\left[\theta(t)\right]R_x^\dagger(\phi)R_y^\dagger\left[\theta(t)\right].
\end{eqnarray}
This choice guarantees that the initial (final) state lying on the periodic Bloch-sphere trajectory shown in Fig.~\ref{fig: trajectory illustration} is correctly mapped from (back to) an instantaneous eigenstate. Substituting the expression for $U_\eta(t)$ from Eq.~\eqref{eq:UetaAppendix} and the expression for $\mathcal{R}(t)$ from Eq.~\eqref{eq:Rt-in-terms-of-S} into the expression for the population transfer error [Eqs.~\eqref{eq: epsilon conversion error} and \eqref{eq:error}], and using $\ket{0(\tf)}=R_y\left[\theta(\tf)\right]\ket{\downarrow}$, $\ket{1(\tf)}=R_y\left[\theta(\tf)\right]\ket{\uparrow}$,  gives a simple expression for the error, independent of the choice of weights $w_j$:
\begin{equation}\label{eq:ErrorGeneralized}
    \epsilon = \left\llangle\left|\bra{\uparrow}\mathcal{T}\exp\left\{-i\int_0^{\tf}dt V_\eta(t)\right\}\ket{\downarrow}\right|^2\right\rrangle_\eta,
\end{equation}
where $\ket{\uparrow},\,\ket{\downarrow}$ are eigenstates of $\sigma_z$. In the absence of noise ($V_\eta(t)=0$), the error vanishes identically for all final times given an ideal generalized fast-QUAD pulse. To account for noise, we expand Eq.~\eqref{eq:ErrorGeneralized} to leading nontrivial order in $\eta$, which recovers the result given in Eq.~\eqref{eq:epsilon-eta-delta} from the main text:
\begin{eqnarray}
    \epsilon & = & \epsilon_\eta(\delta)+\mathcal{O}(\eta^3),\\
    \epsilon_\eta(\delta) & = & \frac{1}{2} \sum_{\alpha,\beta}\int_{-\infty}^\infty \frac{d\omega}{2\pi}\frac{S_{\alpha\beta}(\omega)}{\omega^2}F_{\alpha\beta}(\omega,\tf),
    \label{eq:epsilon-eta-delta-appendix}
\end{eqnarray}
where the noise spectral density is
\begin{equation}
    S_{\alpha\beta}(\omega) = \int_{-\infty}^{\infty}dt\ e^{-i\omega t}\llangle\eta_\alpha(t)\eta_\beta\rrangle_\eta.
\end{equation}
Here, we have introduced the generalized filter function (where the overline indicates complex conjugation):
\begin{equation}
    F_{\alpha\beta}(\omega,\tf) = 2\mathrm{Re}\left[Z_\alpha(\omega,\tf)\overline{Z}_\beta(\omega,\tf)\right],
    \label{eq:appendix-filter-function}
\end{equation}
and where the dimensionless parameters $Z_\alpha$ are given in terms of elements of the rotation matrix $\left[M(t)\right]$ [Eq.~\eqref{eq:Mmatrix}] by
\begin{eqnarray}
    Z_\alpha(\omega,\tf) & = & \omega\int_0^{\tf} dt\ e^{-i\left[\omega t+\Phi_0(t)\right]}M_{\alpha}^\perp(t),\\
    M_{\alpha}^\perp(t) & = & \frac{1}{2}\left[M_{x\alpha}(t)+iM_{y\alpha}(t)\right].
\end{eqnarray}

\subsection{Standard fast-QUAD protocol}
For the standard fast-QUAD protocol, we set the preparation and measurement unitaries to the identity:
\begin{equation}
    \mathcal{R}(t)=\mathbb{1}.
\end{equation}
We then perform a dual expansion in both the amplitude of the noise $\eta$ and in the adiabaticity parameter $\delta$, giving
\begin{equation}
    \epsilon = \tilde{\epsilon}(\delta)+\epsilon_\eta(0)+\mathcal{O}(\eta^2\delta,\eta^3).
\end{equation}
In the adiabatic limit $\delta\to 0$, the contribution from the noise $\epsilon_\eta(0)$ is identical to the previously calculated result [Eq.~\eqref{eq:epsilon-eta-delta-appendix}] for the generalized fast-QUAD protocol.
The exact contribution $\tilde{\epsilon}(\delta)$ due to the coherent dynamics of the state in the absence of noise ($\eta\to 0$) can be found for any $\delta$:
\begin{eqnarray}
    \tilde{\epsilon}(\delta) & = & \left|\bra{\uparrow}R_x(\phi)e^{-\frac{i}{2}\Phi_0(t)\sigma_z}R_x^\dagger(\phi)\ket{\downarrow}\right|^2 \\
    &=& \left[\frac{\dth(\tf)}{2}\right]^2\mathrm{sinc}^2\left[\sqrt{1+\delta^{-2}}\dth(\tf)/2\right],
\end{eqnarray}
which recovers Eq.~\eqref{eq:epsilon0} from the main text.

\section{Filter function for the constant-gap model\label{app: filterfunction}}

In the constant-gap model with polarized noise along the $z$-direction, $\boldsymbol{\eta}(t) = (0,0,\eta(t))^T$, the only nonvanishing component of the spectral density matrix $S_{\alpha\beta}$ is $S_{zz}(\omega)$. In this case, the only relevant component of the filter-function $F_{\alpha\beta}(\omega,t)$ is $F_{zz}(\omega, t)=F(\omega, t)$, Eq.~(\ref{eq:filterfunctionpolarized}): 
\begin{equation}
\label{eq: CG filter function}
F_{zz}(\omega, t_\textrm{f}) =2\left(\frac{\omega\tf}{4}\right)^2 \left|f_+(\omega,\tf) + f_-(\omega,\tf)\right|^2,
\end{equation}
where, given $\phi = \arctan\delta$ and $\delta = \dth(\tf)/\tf B$, we define
\begin{eqnarray}
f_\pm(\omega,\tf) &=& c_\pm e^{-i[\omega+\Delta_\pm(\tf)] \frac{\tf}{2}}\nonumber\\
&& \times\ \sinc\left\{[\omega+\Delta_\pm(\tf)]\frac{\tf}{2}\right\}\\
\Delta_\pm(\tf) &=& (\sqrt{1+\delta^{-2}}\pm1)\delta B,\\
c_\pm &=& -\sin\phi\pm1.
\end{eqnarray}

\section{Noise model\label{subsec: Noise}}

In a realistic system, coupling to an uncontrolled environment can lead to noise sources that may affect the outcome of a state transfer. It is important to include these noise sources to evaluate whether the solution to Eq.~(\ref{eq: diff eq}) is stable under this perturbation. We consider a classical stationary noise source $\eta(t)$.

There are existing strategies to specifically simulate Ornstein-Uhlenbeck noise (stationary Gaussian noise with a Lorentzian spectral density). See, e.g. Ref.~\citep{gillespie1996exact}. In this Appendix we instead describe a more general method to generate any noise with a classical (frequency symmetric) spectrum, also used in Ref.~\citep{yang2019achieving}.

In the weak-noise limit, the population-transfer error can be calculated in terms of the lowest-order nonvanishing (second-order) correlation function $\left<\eta(t)\eta\right>$ for a stationary noise source $\eta(t)$. The second-order correlation function is fully determined by its Fourier transform, the power spectral density $S(\omega)$. The exact spectrum depends on, e.g., the experimental setup, so to characterize/estimate realistic errors, it is important to be able to simulate noise with a desired spectrum. The power spectral density $S(\omega)$ of the noise $\eta(t)$ is given by 
\begin{equation}
S(\omega) = \mathcal{F}\left[ \langle \eta(t) \eta(0) \rangle \right](\omega),
\label{eq: noise spectrum functoin}
\end{equation}
where we have introduced the Fourier transform 
\begin{equation}
\mathcal{F}\left[\langle \eta(t)\eta \rangle\right](\omega) = \int_{-\infty}^\infty dt\,e^{-i\omega t}\langle \eta(t)\eta \rangle     . 
\end{equation}
For a general noise spectrum $S(\omega)$, it may be difficult to sample a realization $\eta(t)$ directly.  However, this can be generated from white noise $w(t)$, where
\begin{equation}
\langle w(t)w(0) \rangle = \kappa\delta(t).
\label{eq: white noise correlation analytic}\\
\end{equation}
Given such a white-noise source, the desired noise $\eta(t)$ can be created by shaping its Fourier spectrum (see Ref.~\onlinecite{smith2011spectral_book}):
\begin{equation}
\eta(t) = \mathcal{F}^{-1}\left[   \sqrt{S(\omega)} \mathcal{F}[w](\omega)   \right](t).
\label{eq: noise shaping}
\end{equation}

\subsection{Discretization of noise \label{sec: discretization of noise}}

\begin{figure}
    \centering
    \includegraphics[width=0.9\columnwidth]{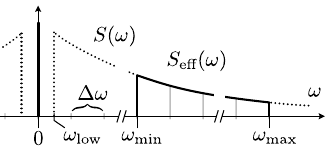}
    \caption{
    Illustration showing the cutoff frequencies of the noise models for $\omega \geq 0$ ($\omega_\tmin$, $\omega_\tmax$), the noise-model spectrum $S(\omega)$ (dotted line) [Eq.~\eqref{eq: 1_f spectrum}]  of the $1/f$-noise model, and the effective noise spectrum $S_\textrm{eff}(\omega)$, Eq.~\eqref{eq: 1_f eff spectrum}, (solid lines) used in the numerical calculations. 
    $\omegaLow$ is the low-frequency cutoff in the $1/f$-noise model, see Eq.~\eqref{eq: 1_f spectrum}. 
    For the numerical calculations, cutoffs $\omega_\tmin$ and $\omega_\tmax$ need to be introduced [Sec.~\ref{sec: simulation cutoffs}]. 
    The part of the spectrum $S(\omega)$ for $\omegaLow < |\omega| < \omega_\tmin$ is collapsed into a zero-frequency [$\propto\delta(\omega)$] component of the effective spectrum $S_\textrm{eff}(\omega)$, illustrated by the vertical line at $\omega=0$. 
    All numerically sampled frequencies are multiples of $\Dom$ [Eq.~\eqref{eq: omega_k}], indicated by the gray tick marks and vertical lines.
    The fast Fourier transform (FFT) requires a number $N$ of datapoints that is a power of two. 
    For that reason, the highest sampling frequency $\omega_\tmax$ is chosen to be $\Dom$, multiplied by a power of two, see Eq.~\eqref{eq: omega_max power of two}.}
    \label{fig:cutoff frequencies}
\end{figure}

In a numerical implementation, white noise $w(t)$ will be created as Gaussian distributed values with mean $\langle w\rangle = 0$ and variance $\langle w^2 \rangle$ in $N$ discrete time-steps $\Dt$ as an array of values
\begin{equation}
w = [w(0),w(\Dt), w(2\Dt),\dots, w(N\Dt)],
\end{equation}
that are (linearly) interpolated to determine the value at a given time $t$. For discrete noise $w$, the correlation function is
\begin{equation}
\langle w(n\Dt)w(0)\rangle
=
\begin{cases}
\langle w^2\rangle & n = 0,\\
0 & \textrm{otherwise}.
\end{cases}\label{eq:discrete noise variance}
\end{equation}
By choosing 
\begin{equation}
\langle w^2 \rangle = \frac{\kappa}{\Dt}
\end{equation}
the desired numerical approximation to the analytical correlation function of the noise, Eq.~(\ref{eq: white noise correlation analytic}), is recovered. 

We use the fast fourier transform (FFT) and its inverse (iFFT) to calculate the frequency spectrum of the discrete noise, apply the desired noise power spectral density $S(\omega)$ and transform back to get the time-series data of the noise, as described leading up to Eq.~\eqref{eq: noise shaping}. The specific implementation of the Java Hipparchus math library \cite{hipparchus} is given by
\begin{eqnarray}
a(\omega_k) &=& \sum_{n=0}^{N-1} \eta(t_n) e^{-i t_n \omega_k} \quad  \text{ forward,} \\
\eta(t_n) &=& \frac{1}{N} \sum_{k=0}^{N-1} a(\omega_k) e^{i t_n \omega_k} \quad \text{inverse,}\\
\omega_k &=& \begin{cases}
k\Dom &\textrm{for }k\leq \frac{N}{2},\\
-(N-k)\Dom &\textrm{for } k \geq \frac{N}{2},\\
\end{cases} \label{eq: omega_k}\\
\Dom &=& 2\pi/N\Dt\\
t_n &=& n\Dt.
\end{eqnarray}

The physical system simulated will require the (dynamical) simulation of a lowest frequency component of the noise $\omega_\tmin$, see Fig.~\ref{fig:cutoff frequencies}. To accurately resolve the noise spectrum around $\omega_\tmin$, the frequency step $\Dom$ has to be smaller, since all sampled frequencies are multiples thereof, so we choose
\begin{equation}
\Dom = 0.1\times \omega_\tmin = 2\pi/t_\tmax,
\label{eq: Dom}
\end{equation}
where, e.g., $t_\tmax = 1~\us$ in the charge-qubit example, Sec.~\ref{sec: charge-qubit}.
Further, to accurately simulate the relevant frequencies of the physical system given by the maximum level splitting $\max_\varepsilon B(\varepsilon)$, the high-frequency cutoff $\omega_\tmax$ must be larger. At the same time, this high-frequency cutoff has to be a power of two multiplied by $\Dom$, as required by the FFT. We therefore choose the next higher power of two:
\begin{eqnarray}
    \omega_\tmax &=& N\Dom/2, \label{eq: omega_max power of two} \\
    N &=&  2^{\lceil\log_2 [ 10\times\max_\varepsilon B(\varepsilon)/\hbar\Dom ] \rceil + 1}, \label{eq: N noise}
\end{eqnarray}
where $\lceil . \rceil$ indicates rounding up to the next integer value.

\subsection{Simulation cutoffs for $1/f$ noise \label{sec: simulation cutoffs}}

The $1/f$-noise model, Eq.~\eqref{eq: 1_f spectrum}, is valid for frequencies $|\omega| \in [\omegaLow, \infty)$. However, a numerical calculation cannot take an infinite range of frequencies into account. Therefore, the two cutoff frequencies explained in Sec.~\ref{sec: discretization of noise} are necessary: a high frequency cutoff $\omega_\tmax$ and a low-frequency cutoff $\omega_\tmin$.

The lowest frequency considered in the $1/f$-noise model $\omegaLow$ is much smaller than the smallest resolved frequency of the noise generation: $\omegaLow \ll \omega_\tmin$. Choosing $\omega_\tmin = \omegaLow$ would, given the resulting small frequency-step size $\Dom$, Eq.~\eqref{eq: Dom}, lead to an impractically large number of datapoints, Eq.~\eqref{eq: N noise}.
Further, since $\Dom$ corresponds to an oscillation period much larger than the population transfer time, see Eq.~\eqref{eq: Dom}, the remaining range of frequencies $\omegaLow < |\omega| < \omega_\tmax$ relevant for the noise simulation contributes approximately only a constant offset to the noise during the population transfer. Therefore, the part of the noise spectrum below the low-frequency cutoff $|\omega|<\omega_\tmin$ will be collapsed into a quasistatic  [$\propto\delta(\omega)$] contribution, giving an effective spectrum
\begin{equation}
S_\textrm{eff}(\omega) = 2\pi\sigma_0^2 \delta(\omega) + \begin{cases}
S(\omega), & \omega_\tmin < |\omega| < \omega_\tmax\\
0, &\textrm{otherwise.}
\end{cases} \label{eq: 1_f eff spectrum}
\end{equation}
The variance $\sigma_0^2$ of the noise for these low frequencies is given by
\begin{equation}
\sigma_0^2 = 2\int_{\omegaLow}^{\omega_\tmin}\frac{d\omega}{2\pi} S(\omega) = \frac{A}{\pi}\ln\round{\frac{\omega_\tmin}{\omegaLow}}.
\end{equation}
With Eq.~\eqref{eq: 1_f eff spectrum}, both the physical spectrum  $S(\omega)$ and the effective spectrum $S_\textrm{eff}(\omega)$, have the same weight for frequencies $|\omega| < \omega_\tmin$
\begin{equation}
\int_{-\omega_\tmin}^{\omega_\tmin} \frac{d\omega}{2\pi} S(\omega) = \int_{-\omega_\tmin}^{\omega_\tmin} \frac{d\omega}{2\pi} S_\textrm{eff}(\omega).
\end{equation}

\subsection{Repository}

The ``Qutlin'' repository of the code used for the numerical simulations can be found at Ref.~\cite{fehse2021Qutlin}.

\bibliography{refs.bib}

\end{document}